# Fabrication of Dense Ultrafine-Grained MoW, MoWNb, and MoWNbTa Alloys: Influence of Cobalt Doping on Sintering and Grain Growth

Keyu Cao [a,†], Sashank Shivakumar [b,†], Jian Luo [a,b,*]

[a] Aiiso Yufeng Li Family Department of Chemical and Nano Engineering, University of California San Diego, La Jolla, California 92093, U.S.A.

[b] Program in Materials Science and Engineering, University of California San Diego, La Jolla 92093, U.S.A.

## Abstract

Dense ultrafine-grained (UFG) refractory MoW, MoWNb, and MoWNbTa alloys were fabricated by combining high-energy ball milling (HEBM) and spark plasma sintering (SPS), achieving ~92–96% relative densities and ~70–180 nm grain sizes. The effects of 2 at.% cobalt (Co) addition on sintering behavior and high-temperature grain growth resistance were investigated as a function of compositional complexity. Activated sintering was observed, with 2 at.% Co addition increasing relative densities from ~92–96% to ~96–98%. Isothermal grain growth experiments at 1200 °C and 1300 °C showed that Co doping suppressed the relative grain growth rate, despite a modest initial grain size increase due to Co-activated sintering, with the effect becoming more pronounced in compositionally complex alloys. The observed trend is consistent with the recently proposed high-entropy grain boundary (HEGB) effect. Notably, $Mo_{24.5}W_{24.5}Nb_{24.5}Ta_{24.5}Co_2$ achieved a 96.4% relative density and maintained an ultrafine grain size, increasing only slightly from ~122 nm to ~127 nm after 5 h annealing at 1200 °C. Scanning transmission electron microscopy (STEM and energy-dispersive X-ray spectroscopy (EDS) confirmed strong Co segregation at grain boundaries, accompanied by minor depletion of Ta and W, supporting a recently proposed grain boundary segregation model for high-entropy alloys and HEGBs.

---

† These authors contribute equally.

* Corresponding Author. Email: jluo@alum.mit.edu (Jian Luo)

# 1. Introduction

Complex concentrated (or compositionally complex) alloys (CCAs), also referred to as high-entropy alloys (HEAs), represent a new class of materials without a single dominant element [1]. In pursuit of advanced high-temperature materials, refractory CCAs (RCCAs), such as MoWNbTa, were introduced in 2010 [2] and have since attracted considerable attention for their promising high-temperature mechanical properties [3–8]. However, the fabrication of RCCAs remains challenging due to their extremely high melting points. In particular, dense ultrafine-grained (UFG) RCCAs, which may exhibit unique mechanical properties, have been difficult to produce, thereby remaining largely underexplored. Furthermore, to enable practical applications, UFG RCCAs must demonstrate sufficient stability against grain growth at elevated temperatures, which have not been demonstrated.

Solid-state activated sintering has been examined to facilitate the densification of refractory metals, where the addition of small amounts of transition metals such as Ni, Co, Fe and Pd promotes sintering of W, Mo, and Ta [9–12]. Prior studies have attributed Ni-activated sintering of W and Mo to enhanced diffusion through liquid-like interfacial phases stabilized at grain boundaries (GBs) below the bulk eutectic or solidus temperatures [10,13,14], leading to the development of GB phase ($\lambda$) diagrams to predict high-temperature GB disordering and related sintering behavior [15–19]. Such computed GB phase ($\lambda$) diagrams have since been extended to ternary alloys [20,21], and mostly recently, to a RCCA, Ni-doped MoWNbTa [22], to predict activated sintering. However, achieving enhanced densification while simultaneously suppressing grain growth remains a major challenge.

The stabilization of nanocrystalline and UFG alloys is generally attributed to both kinetic and thermodynamic stabilization mechanisms [23–25]. In 2016, Zhou et al. introduced the concept of high-entropy grain boundaries (HEGBs) [26], which are used to stabilize nanocrystalline alloys at elevated temperatures [21,26,27]. The concept of HEGBs was further elaborated in a recent perspective article [28]. A thermodynamic model was developed to predict GB segregation in HEAs (CCAs) and to quantify the HEGB effects [29].

In this work, dense UFG refractory MoW, MoWNb, and MoWNbTa alloys were successfully fabricated by combining high-energy ball milling (HEBM) and spark plasma sintering (SPS), achieving >92% relative densities and grain sizes of <200 nm. Activated sintering was observed upon 2 at.% Co addition, which increased the relative densities from ~92–96% to ~96–98%. Isothermal grain growth experiments at 1200 °C and 1300 °C revealed that Co doping suppressed the relative grain growth rate, with the effect becoming more pronounced in CCAs, consistent with the recently proposed Type II HEGB effect [28]. Pronounced Co segregation at GBs, accompanied by minor Ta and W depletion, further supports the recently developed GB segregation model for HEAs/CCAs and HEGBs [29].

## 2. Experimental Procedures

UFG RCCA samples were fabricated via a powder metallurgical route. Elemental powders of Nb (1-5 μm, 99.8 %), Mo (3-7 μm, 99.95 %), Ta (2 μm, 99.9 %), W (44 μm, 99.9 %), and Co (37 μm, 99.5 %) were all purchased from Thermo Fisher Scientific (MA, USA). The powders were weighed in appropriate proportions and vortex-mixed for 5 min. Steric acid (0.1 wt.%) was added as a process control agent (PCA) to prevent excessive cold welding and agglomeration, while producing minor (<2-3 vol.%) secondary carbide phases, following a recent study [30].

The mixed powders were mechanically alloyed in 6 g batches using a SPEX 8000D high-energy ball mill (HEBM, Cole-Parmer SamplePrep, USA) for 6 h, employing WC-lined stainless-steel jars and 11.2 mm diameter WC balls at a ball-to-powder ratio of ~4:1. All powder handling and jar sealing were carried out in an Ar-filled glovebox (<10 ppm $O_2$). Milling was performed in 30 min intervals with 10 min cooling periods. Prior to each milling run, 2 g of sacrificial powder with the same composition was milled for 2 h to precoat the milling media, thereby minimizing contamination and improving powder yield. Mechanically alloyed powders were loaded in ~ 2 g batches into 10 mm graphite dies (California Nanotechnologies, CA, USA) lined with graphite and tantalum foil, and subsequently consolidated by spark plasma sintering (SPS) under vacuum ($10^{-2} – 10^{-3}$ torr) using a Thermal Technologies 3000 Series SPS system. SPS was carried out at 1200 °C under a uniaxial pressure of 50 MPa, employing a heating rate of 100 °C/min and a dwell time of 5 min. Samples were furnace cooled under vacuum following SPS. The as-sintered pellets were manually ground with SiC grinding paper to remove foil remnants and any surface contamination layers.

Densities were measured using the Archimedes method, with theoretical densities calculated based on the rule of mixtures. Isothermal annealing was performed at 1200 °C and 1300 °C for 5 h in a tube furnace under flowing Ar + 5% H$_2$. Titanium sponge (3–19 mm, 99.95 %, ThermoFisher Scientific) was placed at the tube inlet as an oxygen getter. A heating rate of 10 ºC/min to the targeted isothermal annealing temperature was employed, followed by furnace cooling after completion of the isothermal annealing.

X-ray diffraction (XRD) was performed on the milled powders, as well as as-sintered and annealed pellets, using a Rigaku Miniflex diffractometer with Cu K$\alpha$ radiation (30 kV, 15mA). Samples were fractured after embrittlement in liquid nitrogen (LN$_2$). The resulting fracture surfaces were examined using scanning electron microscopy (SEM). Grain size measurements were obtained from corresponding SEM images of fracture surfaces in accordance with ASTM E112-88.

Samples were also hot mounted in epoxy, ground and polished to a mirror finish. Energy dispersive X-Ray spectroscopy (EDS) was performed on a Thermo Fisher Apreo SEM equipped with an Oxford N-Max EDS detector.

For transmission electron microscopy (TEM), electron-transparent lamellas were first produced via focused ion beam (FIB) milling using a Thermo Fisher Scios DualBeam FIB/SEM. Scanning transmission electron microscopy (STEM) with EDS elemental mapping was conducted using a Thermo Fisher Talos F200X TEM.

## 3. Results and Discussion

### 3.1. As-sintered UFG RCCAs and Phase Formation

Table 1 summarizes the sample compositions, sintering temperatures, as-sintered densities, and corresponding grain sizes. All six as-sintered specimens attained relative densities of ~92.5–98% and ultrafine grain sizes of ~70–190 nm. As-sintered Mo$_{50}$W$_{50}$ and Mo$_{49}$W$_{49}$Co$_2$ achieved the highest relative densities of 96.6% and 98.4%, respectively, whereas the other compositionally complex alloys exhibited lower densities (~92–94%) without Co doping, which increased to ~96% with Co doping.

Figure 1 shows XRD patterns of all as-milled powders and SPS-sintered samples. After 6 h of HEBM, all powders formed a single BCC phase with minor metal carbide secondary phases (indexed as M$_2$C). The primary BCC peaks gradually shifted to lower angles from Mo$_{50}$W$_{50}$ to Mo$_{25}$W$_{25}$Nb$_{25}$Ta$_{25}$ due to the addition of Nb and Ta, which have larger lattice parameters (~0.146 nm) compared to Mo and W (~0.139 nm). The carbide secondary phases were introduced by the 0.1 wt.% PCA used during powder processing [30]. The as-sintered specimens exhibited similar phase compositions as the as-milled powders, with a few compositions showing minor additional MC phases. The XRD peaks of the sintered samples are sharper, indicating improved crystallinity and reduced microstrain. Any Co-based secondary phases, if present, are below the XRD detection limit.

### 3.2. Activated Sintering

As shown in Table 1, the relative densities of as-sintered $Mo_{50}W_{50}$, $Mo_{33.3}W_{33.3}Nb_{33.3}$, $Mo_{25}W_{25}Nb_{25}Ta_{25}$ were measured as 96.6%, 92.5%, and 93.6%, respectively. Upon the addition of 2 at.% Co, the relative densities of as-sintered $Mo_{49}W_{49}Co_2$, $Mo_{32.7}W_{32.7}Nb_{32.6}Co_2$, and $Mo_{24.5}W_{24.5}Nb_{24.5}Ta_{24.5}Co_2$ increased to 98.4%, 96%, and 96.4%, respectively. The observed ~2–4% increase in relative density with Co addition can be attributed to activated sintering, here facilitated under SPS conditions rather than conventional sintering.

SEM images of the fractured surfaces for all specimens are shown in Figure 2. The mean grain sizes were measured as $178 \pm 6.6$ nm for $Mo_{50}W_{50}$, $190 \pm 32$ nm for $Mo_{49}W_{49}Co_2$, $68.7 \pm 25.6$ nm for $Mo_{33.3}W_{33.3}Nb_{33.3}$, $102.8 \pm 5.8$ nm for $Mo_{32.7}W_{32.7}Nb_{32.6}Co_2$, $94.5 \pm 7.2$ nm for $Mo_{25}W_{25}Nb_{25}Ta_{25}$, and $122.8 \pm 8.3$ nm for $Mo_{24.5}W_{24.5}Nb_{24.5}Ta_{24.5}Co_2$. The Co-doped specimens, which are ~2-4% denser, exhibited slightly larger grain sizes compared to their undoped counterparts, consistent with the positive correlation between grain size and relative density [10,27]. It is also possible that Co addition promotes both densification and grain coarsening through enhanced grain boundary diffusion [31], although this effect is difficult to isolate and likely minor, given that the overall grain size increase is moderate and may primarily reflect reduced porosity.

The relationship between grain size and relative density for all samples is plotted in Figure 3, showing that 2 at.% Co addition substantially enhanced densification (reducing porosity by ~43–52%) with only moderately increasing grain growth (increasing the grain size by ~12-36 nm or ~7-54% from the undoped specimens).

### 3.3. Grain Growth Experiment and the Effect of Co Doping

Isothermal grain growth experiments were conducted on all specimens at 1200 °C and 1300 °C, respectively, for 5 h. XRD patterns of annealed specimens (Supplementary Fig S1) show no significant changes after annealing, as expected. The primary BCC phase was retained after 5 h anneal, while the secondary metal carbide phases remained detectable.

SEM–EDS mapping was performed for all compositions in both as-sintered and 1200 °C annealed samples, as shown in Supplementary Fig S2 and Fig S3. The as-sintered samples exhibited relatively homogeneous bulk compositions, with minor inhomogeneities in Co and W that can be attributed to the short SPS dwell time and resulting nonequilibrium conditions. After annealing at 1200 °C for 5 h, all elements were uniformly distributed across all systems. The only isolated W-rich region observed in $Mo_{25}W_{25}Nb_{25}Ta_{25}$ is likely a contaminant introduced from the WC milling jar during the HEBM process.

The SEM images of annealed specimens are shown in Supplementary Fig S4. After annealing at 1200 °C for 5 h, mean grain sizes were measured to be 313.8 ± 14.1 nm for $Mo_{50}W_{50}$, 297.9 ± 27.2 nm for $Mo_{49}W_{49}Co_2$, 142.7 ± 13.2 nm for $Mo_{33.3}W_{33.3}Nb_{33.3}$, 163.2 ± 13.9 nm for $Mo_{32.7}W_{32.7}Nb_{32.6}Co_2$, 149.3 ± 5.9 nm for $Mo_{25}W_{25}Nb_{25}Ta_{25}$, and 127.9 ± 7.8 nm for $Mo_{24.5}W_{24.5}Nb_{24.5}Ta_{24.5}Co_2$. After annealing at 1300 °C for 5 h, the corresponding mean grain sizes increased to 333.7 ± 24 nm, 381.7 ± 25.5 nm, 231.6 ± 22.1 nm, 233.6 ± 5.7 nm, 218.1 ± 8.9 nm, and 189.7 ± 5.8 nm, respectively. The measured grain sizes of all annealed samples were summarized in Table 2.

The percentages of grain size change (increase) after 5 h of isothermal annealing (referred to as "grain growth percentage") were also calculated and summarized in Table 2, and are plotted in Figure 4(a). Compared with the undoped specimens, the addition of 2 at.% Co reduced the grain growth percentages in most cases (except for MoW at 1300 °C), as shown in Figure 4(a). For annealing at 1200 °C, the grain growth percentage decreased from ~76 % $Mo_{50}W_{50}$ to ~57% for $Mo_{49}W_{49}Co_2$, and from ~108 % for $Mo_{33.3}W_{33.3}Nb_{33.3}$ to ~59% for $Mo_{32.7}W_{32.7}Nb_{32.6}Co_2$, and from ~58% for $Mo_{25}W_{25}Nb_{25}Ta_{25}$ to ~4% for $Mo_{24.5}W_{24.5}Nb_{24.5}Ta_{24.5}Co_2$. As shown in Figure 4(b), the addition of 2 at.% Co reduced the grain growth percentage by ~26% for the MoW-based alloys (two principal elements), by ~45% for the MoWNb-based alloys (three principal elements), and by as much as ~93% for the MoWNbTa-based alloys (four principal elements).

For annealing at 1300 °C, the grain growth percentage increased from ~88 % $Mo_{50}W_{50}$ to ~101% for $Mo_{49}W_{49}Co_2$, while it decreased from ~237 % for $Mo_{33.3}W_{33.3}Nb_{33.3}$ to ~127% for $Mo_{32.7}W_{32.7}Nb_{32.6}Co_2$, and from ~131% for $Mo_{25}W_{25}Nb_{25}Ta_{25}$ to ~55% for $Mo_{24.5}W_{24.5}Nb_{24.5}Ta_{24.5}Co_2$. As shown in Figure 4(c), the addition of 2 at.% Co increased the grain growth percentage by ~15% for the MoW-based alloys (two principal elements), but it reduced the grain growth percentage by ~46% for the MoWNb-based alloys (three principal elements) and by ~58% for the MoWNbTa-based alloys (four principal elements).

This observed trend, showing increasing stabilization effects from Co doping (and associated GB segregation, as shown later in Section E) with increasing number of principal elements, can be attributed to the recently proposed Type II HEGB effect [28,29], which will be elaborated in Section D.

The stabilization effect with Co doping is most pronounced in $Mo_{24.5}W_{24.5}Nb_{24.5}Ta_{24.5}Co_2$, which exhibited only ~4% and ~54% grain growth—from ~123 nm to ~128 nm and ~190 nm—after annealing at 1200 °C and 1300 °C for 5 h, respectively. In comparison, the undoped $Mo_{25}W_{25}Nb_{25}Ta_{25}$ experienced ~58 % and ~130 % grain growth from ~95 nm to ~150 nm and ~218 nm under the same conditions. The corresponding SEM images are shown in Figure 5. Notably, despite having a higher relative density and larger initial grain size after SPS, the Co-doped $Mo_{24.5}W_{24.5}Nb_{24.5}Ta_{24.5}Co_2$ maintained a smaller grain size after high-temperature annealing compared with the undoped $Mo_{25}W_{25}Nb_{25}Ta_{25}$.

Figure 6 shows TEM bright-field images taken on as-sintered and 1200 °C-annealed $Mo_{25}W_{25}Nb_{25}Ta_{25}$ and $Mo_{24.5}W_{24.5}Nb_{24.5}Ta_{24.5}Co_2$, further confirming grain-size measurements obtained by SEM. The grain sizes after sintering, measured from TEM images, were 102.9 ± 42.9 nm for $Mo_{25}W_{25}Nb_{25}Ta_{25}$, and 121.9 ± 41.6 nm for $Mo_{24.5}W_{24.5}Nb_{24.5}Ta_{24.5}Co_2$. After annealing at 1200 °C for 5 h, the TEM-measured grain sizes increased to 164.1 ± 64.4 nm and 134.3 ± 58.1 nm, respectively. The TEM-measured grain sizes show good agreement with those measured by SEM. A comparison of the SEM- and TEM-measured grain sizes is provided in Table S1.

**3.4. Thermodynamic Modeling** In general, activated sintering promotes both densification and grain coarsening by enhancing diffusion along GBs. In the present study, Co addition moderately increased the grain size while simultaneously enhancing densification through activated sintering. Interestingly, however, Co addition also improved the stability against grain growth during subsequent high-temperature annealing, which may be attributed to thermodynamic stabilization effects [23,32,33].

An interesting observation in this study is that the stabilization effect induced by Co doping becomes more pronounced with increasing compositional complexity, which may be attributed to the recently proposed Type II HEGB effect. This effect can be illustrated using an analytical model recently developed by Luo [29] for GB segregation in a multicomponent ideal solution (particularly applicable to HEAs and CCAs), in which GB segregation enthalpies are estimated from density functional theory (DFT) data available in the Materials Project [34]. In this model [29], the GB composition of element $i$ is derived as

$$X_i^{GB} = \frac{X_i^{Bulk} e^{-\frac{\Delta h_{ads.(i)}^{(0)}}{kT}}}{\sum_j X_j^{Bulk} e^{-\frac{\Delta h_{ads.(j)}^{(0)}}{kT}}} \tag{1}$$

where $k$ is the Boltzmann constant, and $X_i^{GB}$ and $X_i^{Bulk}$ are the atomic fractions of element $i$ at the GB and in the bulk, respectively. The GB adsorption (segregation) enthalpy of element $i$, $\Delta h_{ads.(i)}^{(0)}$, can be expressed as

$$\Delta h_{ads.(i)}^{(0)} = \tfrac{1}{2} Q z_v \left( |e_{ii}| - \left| \sum_j X_j^{Bulk} \cdot e_{jj} \right| \right) - \Delta E_i^{strain} \tag{2}$$

where $z_v$ is the number of bonds between two neighboring atomic layers parallel to the twist GB, and $Q$ is the fraction of bonds broken at the GB core between the two GB planes (set to 1/6 to represent an average large-angle general GB, consistent with the empirical relation $\gamma_{GB}/\gamma_{surface} = 1/3$). $e_{ii}$ (or $e_{jj}$) is the self-bonding energy of element $i$ (estimated from DFT data), and $\Delta E_{strain}^i$ is the elastic strain energy of element $i$ in the bulk.

In this analytical model [29], the GB energy was derived as

$$\gamma_{GB} = \overline{\gamma_{GB}^{(0)}} - \Gamma_0 kT \ln\left( \sum_i X_i^{Bulk} e^{-\frac{\Delta h_{ads.(i)}^{(0)}}{kT}} \right) \tag{3}$$

where $\Gamma_0$ is the number of GB sites per unit area. Here,

$$\overline{\gamma_{GB}^{(0)}} = \sum_j X_j^{Bulk} \cdot \gamma_{GB,j}^{(0)} \tag{4}$$

and

$$\gamma_{GB,j}^{(0)} = \tfrac{1}{2}\Gamma_0 Q z_v |e_{jj}| \tag{5}$$

represents the contribution of broken bonds from all components to the GB energy, without considering adsorption or relaxation effects. A detailed derivation of this analytical model is provided in the prior publication [29]. For a Co-saturated multicomponent alloy with $(N-1)$ principal elements (plus Co as the component $N$), the bulk solid solubility of Co can be estimated, following the prior publication [29], as:

$$X_N^{Bulk} \approx \min_{i=1,2,..(N-1)} \left[ \left(X_i^{Bulk}\right)^{-\beta_{N(i)}} X_{N(i)}^{Binary\ Solvus} \right] \tag{6}$$

where $X_{N(i)}^{Binary\ Solvus}$ is the Co solubility along the binary solvus line for the $i$–$N$ ($N$ = Co) system in equilibrium with a precipitate of the binary $M_{x_i}Co_{y_i}$ line compound, and $\beta_{N(i)} = x_i / y_i$.

This thermodynamic model is applied to the RCCAs. The bulk solubilities of Co in each RCCAs were calculated by Eq. (6). The binary solvus line data of M-Co (M = Mo, W, Nb, Ta) was extracted from the phase diagrams calculated by CALPHAD, as shown in Fig S5. The temperature-dependent bulk compositions of the Co-saturated alloys were determined through an iterative process that accounted for mass balance after precipitation, which slightly altered the ratios of the principal elements in the BCC phase (since only 2 at.% Co was added). All other parameters used in the model, such as atomic radii and shear moduli, were obtained from DFT data available in the Materials Project and are summarized in Supplementary Table S3.

Figure 7 shows the calculated GB energies as a function of temperature for undoped and Co-saturated MoW, MoWNb, and MoWNbTa alloys. In the undoped alloys, GB energies increase with temperature, which can be attributed to temperature-induced desorption. In contrast, the GB energies of Co-saturated alloys decrease with temperature, indicating a positive effective GB entropy. Moreover, the reduction in GB energy in Co-saturated RCCAs becomes more pronounced with increasing compositional complexity, from MoW to MoWNb and then to MoWNbTa. This behavior represents a Type II HEGB effect [28,29], lowering the GB energy and thereby reducing the driving force for grain growth. This thermodynamic stabilization, which is moderate in the current system (Figure 7), aligns with experimental observations: 2 at.% Co addition decreases grain growth percentages, with the effect becoming more significant as the number of principal elements increases (MoWNbTa > MoWNb > MoW, Figure 4).

This stabilization effect is likely caused by Co segregation at grain boundaries (GBs), as confirmed by STEM-EDS in the next section. Co segregation can have multiple impacts. In addition to thermodynamic stabilization via reduced GB energy (moderate in these systems, as shown in Figure 7), it can provide kinetic stabilization through solute drag, though this can be partially offset by enhanced GB diffusion induced by Co, as supported by activated sintering. Co segregation can also lead to GB disordering, potentially forming liquid-like interfacial phases, which has two competing effects on grain growth. On the one hand, GB disordering increases GB entropy and attracts further adsorption, enhancing both thermodynamic and kinetic stabilization (e.g., the reduction in GB energy in a disordered GB could exceed that predicted in Figure 7). On the other hand, disordered (liquid-like) GBs can significantly enhance GB diffusion, promoting grain growth. In this study, the interplay of these effects results in a moderate reduction in grain growth rates, which becomes more pronounced with increasing compositional complexity, consistent with the proposed Type II HEGB effect [28,29].

### 3.5. GB Composition: Thermodynamic Modeling *vs.* STEM-EDS

Furthermore, we calculated the GB composition for the best-performing sample, $Mo_{24.5}W_{24.5}Nb_{24.5}Ta_{24.5}Co_2$ (*i.e.*, Co-saturated MoWNbTa for the bulk BCC phase). As shown in Figure 8(b), the model predicts strong GB segregation of Co, accompanied by moderate depletion of W and Ta. This pronounced Co segregation at GBs is the key mechanism underlying the observed grain-growth suppression, consistent with the proposed Type II HEGB effect.

To verify the model predictions and further evaluate the GB segregation model and associated HEGB theory, STEM imaging and EDS elemental mapping were performed on $Mo_{24.5}W_{24.5}Nb_{24.5}Ta_{24.5}Co_2$. Nanoscale EDS maps of all constituent elements were acquired from both the bulk and GB regions, as shown in Figure 8(a). A pronounced Co segregation, accompanied by minor W and Ta depletion along the GB, was clearly observed, consistent with the model prediction in Figure 8(b).

Figure 8(c) further shows the line compositional profiles of the five elements across the GB, normalized to their respective bulk compositions ($X_i^{\text{GB}}/X_i^{\text{Bulk}}$). The profiles clearly reveal strong Co segregation accompanied by moderate Ta and W depletions, consistent with the model prediction in Figure 8(b). Thus, these STEM-EDS observations validate the GB segregation model and support the HEGB theory, wherein Co segregation suppresses grain growth at high temperatures, even though it also promotes activated sintering.

In addition, the metal carbide phase observed in XRD was confirmed to be Nb- and Ta-rich, as shown in Supplementary Figure S6. The enrichment of Nb and Ta in the carbide phase is expected, given their higher carbide formation energies, which make Nb and Ta energetically favorable to preferentially dissolve into the carbide phase [30].

## 4. Conclusions

In summary, this study fabricated a series of UFG RCCAs with increasing compositional complexity, namely, undoped and 2 at.% Co-doped MoW, MoWNb, and MoWNbTa, by combining HEBM and SPS. The resulting dense, bulk UFG RCCAs achieved relative densities of ~92–98% and grain sizes of ~70–180 nm.

We systematically examined the effects of 2 at.% Co addition on sintering behavior and grain growth. Under SPS conditions, Co doping promoted activated sintering, increasing the relative density from ~92–96% to ~96–98% with moderate grain growth. Isothermal grain growth experiments conducted at 1200 °C and 1300 °C further revealed that Co addition suppressed the relative grain growth rate, with the effect becoming more pronounced as the compositional complexity increased, consistent with the recently proposed HEGB effect [28,29]. Notably, the $Mo_{24.5}W_{24.5}Nb_{24.5}Ta_{24.5}Co_2$ alloy achieved a 97.6% relative density and retained an ultrafine grain size, increasing only slightly from ~122 nm to ~127 nm after 5 h of annealing at 1200 °C.

To rationalize the observed Co effect on grain growth suppression, we applied a recently developed GB segregation model for HEAs (CCAs) [29] to calculate GB energies and GB composition. The model predicted strong Co segregation at GBs accompanied by moderate Ta and W depletion. STEM and EDS elemental mapping confirmed this prediction, revealing pronounced Co segregation and minor Ta and W depletion along GBs. These results validate the recently developed GB segregation model for HEAs (CCAs) [29] and provide experimental support for the proposed HEGB theory [28,29].

## CRediT authorship contribution statement

**Keyu Cao:** Data curation, Formal analysis, Writing – Original draft, Investigation, Validation. **Sashank Shivakumar:** Writing – Review & Editing, Data curation, Investigation. **Jian Luo:** Supervision, Writing – Review & Editing, Funding acquisition, Conceptualization.

## Declaration of competing interest

The authors declare that there are no known conflicts or competing interest that could have appeared to influence the work reported in this paper.

## Acknowledgement

This work was supported by the U.S. Army Research Office (ARO) under Grant No. W911NF2210071, through the Synthesis & Processing Science of Extreme Materials program. We thank our ARO program manager, Dr. Jacob C. Marx, for his guidance and support.

## Appendix A. Supplementary data

Supplementary data related to this article can be found online at

## Data availability

Data will be made available on request.

**List of Figure Captions**

**Figure 1** X-ray diffraction (XRD) patterns of (a) powders after high-energy ball milling (HEBM) for 6 h and (b) pellets after spark plasma sintering (SPS) at 1200 °C and 50 MPa. All specimens exhibit a BCC primary phase, along with minor secondary metal carbide phases attributed to the process control agent (~0.1 wt.% steric acid) during HEBM as a carbon source.

**Figure 2** Scanning electron microscopy (SEM) images of fractured surfaces of as-sintered specimens (after SPS at 1200 °C and 50 MPa for 5 min), showing the effects of 2 at.% Co addition. The measured relative densities and average grain sizes are indicated.

**Figure 3** Grain size as a function of relative density for as-sintered specimens of six compositions: equimolar MoW, MoWNb, and MoWNbTa, each without and with 2 at.% Co addition. The Co addition increased the relative densities by approximately 2–4 %, accompanied by moderate increases in grain sizes in the sintered specimens, likely due to both enhanced GB diffusion and reduced porosity.

**Figure 4** (a) Grain growth percentages for all six compositions after isothermal annealing at 1200 °C and 1300 °C for 5 h. Results for undoped specimens are shown by open circles while results for specimens with 2 at.% Co addition are shown by the solid circles and labeled as "+2% Co," and the changes upon Co addition are indicated by arrows. Reduction in grain growth percentage upon 2 at.% Co addition for annealing at (b) 1200 °C and (c) 1300 °C. The results indicate that Co addition more effectively suppresses grain growth with increasing number of principal elements, following the order MoW → MoWNb → MoWNbTa, consistent with the hypothesized Type II HEGB effect [28,29].

**Figure 5** Scanning electron microscopy (SEM) images of fractured surfaces of (a, b) $Mo_{25}W_{25}Nb_{25}Ta_{25}$ (*i.e.*, MoNbTaW) and (c, d) $Mo_{24.5}W_{24.5}Nb_{24.5}Ta_{24.5}Co_2$ (*i.e.*, MoNbTaW + 2 at.% Co) specimens after annealing at (a, c) 1200 °C and (b, d) 1300 °C for 5 h, respectively. The corresponding as-sintered microstructures are shown in Figure 2. The addition of Co reduced the extent of grain growth after annealing, despite the higher densities of the doped specimens. The percentage increases in grain size after 5 h of isothermal annealing are indicated for each case.

**Figure 6** Transmission electron microscopy (TEM) bright-field images of (a, b) $Mo_{25}W_{25}Nb_{25}Ta_{25}$ (*i.e.*, MoNbTaW) and (c, d) $Mo_{24.5}W_{24.5}Nb_{24.5}Ta_{24.5}Co_2$ (*i.e.*, MoNbTaW + 2 at.% Co) specimens before (a, c) and after (b, d) annealing at 1200 °C for 5 h. The SEM- and TEM-measured grain sizes are compared in Supplementary Table S1, which are consistent each other.

**Figure 7** Computed GB energy ($\gamma_{GB}$) as a function of temperature for undoped (dashed lines) and Co-saturated (solid lines) MoW, MoWNb, and MoWNbTa alloys. The bulk compositions of the Co-saturated BCC alloys are temperature dependent and follow the multicomponent solvus lines estimated by Equation (6), assuming a 2 at.% Co addition in equimolar RCCAs. The calculations were performed using an analytical model recently proposed by Luo [29], with model parameters obtained from the Materials Project (Supplementary Table S3).

**Figure 8** (a) STEM image and EDS elemental maps of $Mo_{24.5}W_{24.5}Nb_{24.5}Ta_{24.5}Co_2$ after annealing at 1200 °C for 5 h. (b) Grain-boundary (GB) atomic fraction of each element predicted by the simplified segregation model. (c) Normalized EDS elemental intensity at the GB ($X_i^{GB}/X_i^{Bulk}$) as a function of position, obtained by integrating the EDS maps along the direction perpendicular to the GB plane located at x = 0. The data reveal strong Co segregation and mild Ta and W depletion at the GB, in agreement with the model prediction shown in panel (b).

**Table 1** Summary of relative densities and average grain sizes of all as-sintered specimens. All samples were fabricated by spark plasma sintering (SPS) at 1200 °C and 50 MPa for 5 min. Theoretical densities were calculated using the rule of mixtures.

| Composition | Theoretical Density (g/cm$^3$) | Density (g/cm$^3$) | Relative Density | Grain Size (nm) |
|---|---|---|---|---|
| $Mo_{50}W_{50}$ | 14.82 | 14.31 | 96.6% | 178 ± 6.6 |
| $Mo_{49}W_{49}Co_2$ | 14.73 | 14.49 | 98.4% | 190 ± 32 |
| $Mo_{33.3}W_{33.3}Nb_{33.3}$ | 12.54 | 11.6 | 92.5% | 68.7 ± 25.6 |
| $Mo_{32.7}W_{32.7}Nb_{32.6}Co_2$ | 12.49 | 11.99 | 96.0% | 102.8 ± 5.8 |
| $Mo_{25}W_{25}Nb_{25}Ta_{25}$ | 13.65 | 12.78 | 93.6% | 94.5 ± 7.2 |
| $Mo_{24.5}W_{24.5}Nb_{24.5}Ta_{24.5}Co_2$ | 13.59 | 13.1 | 96.4% | 122.8 ± 8.3 |

Table 2 Summary of the key results of the grain growth experiments after isothermal annealing in Ar + 5% $H_2$ for 5 h.

| Composition | Annealing Temperature (°C) | Grain Size (nm) | Grain Size Change after Annealing |
|---|---|---|---|
| $Mo_{50}W_{50}$ | 1200 | 313.8 ± 14.1 | 76.3% |
| | 1300 | 333.7 ± 24 | 87.5% |
| $Mo_{49}W_{49}Co_2$ | 1200 | 297.9 ± 27.2 | 56.8% |
| | 1300 | 381.7 ± 25.5 | 101% |
| $Mo_{33.3}W_{33.3}Nb_{33.3}$ | 1200 | 142.7 ± 13.2 | 107.7% |
| | 1300 | 231.6 ± 22.1 | 237% |
| $Mo_{32.7}W_{32.7}Nb_{32.6}Co_2$ | 1200 | 163.2 ± 13.9 | 58.8% |
| | 1300 | 233.6 ± 5.7 | 127.2% |
| $Mo_{25}W_{25}Nb_{25}Ta_{25}$ | 1200 | 149.3 ± 5.9 | 58.0% |
| | 1300 | 218.1 ± 8.9 | 130.8% |
| $Mo_{24.5}W_{24.5}Nb_{24.5}Ta_{24.5}Co_2$ | 1200 | 127.9 ± 7.8 | 4.1% |
| | 1300 | 189.7 ± 5.8 | 54.5% |

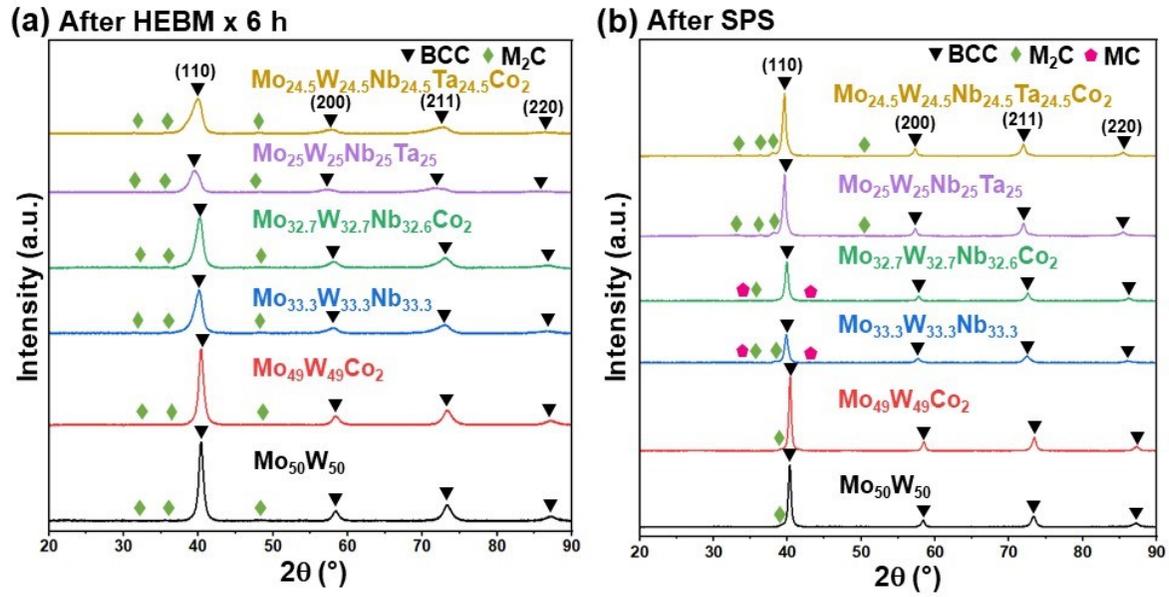

**Figure 1** X-ray diffraction (XRD) patterns of (a) powders after high-energy ball milling (HEBM) for 6 h and (b) pellets after spark plasma sintering (SPS) at 1200 °C and 50 MPa. All specimens exhibit a BCC primary phase, along with minor secondary metal carbide phases attributed to the process control agent (~0.1 wt.% steric acid) during HEBM as a carbon source.

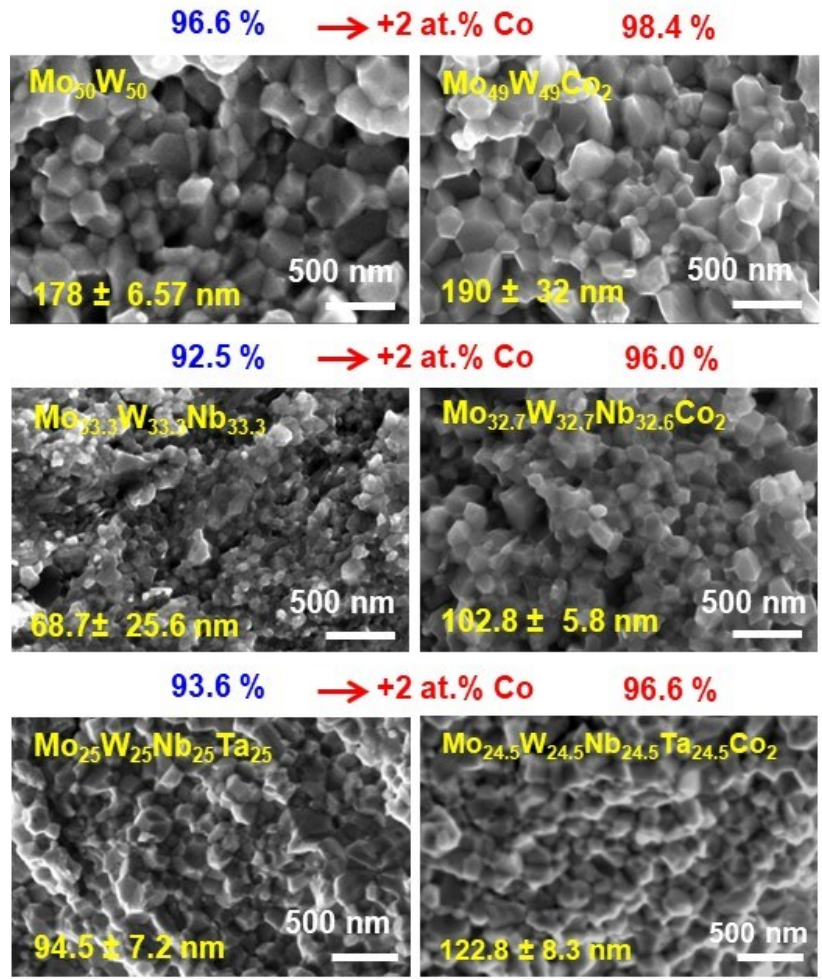

**Figure 2** Scanning electron microscopy (SEM) images of fractured surfaces of as-sintered specimens (after SPS at 1200 °C and 50 MPa for 5 min), showing the effects of 2 at.% Co addition. The measured relative densities and average grain sizes are indicated.

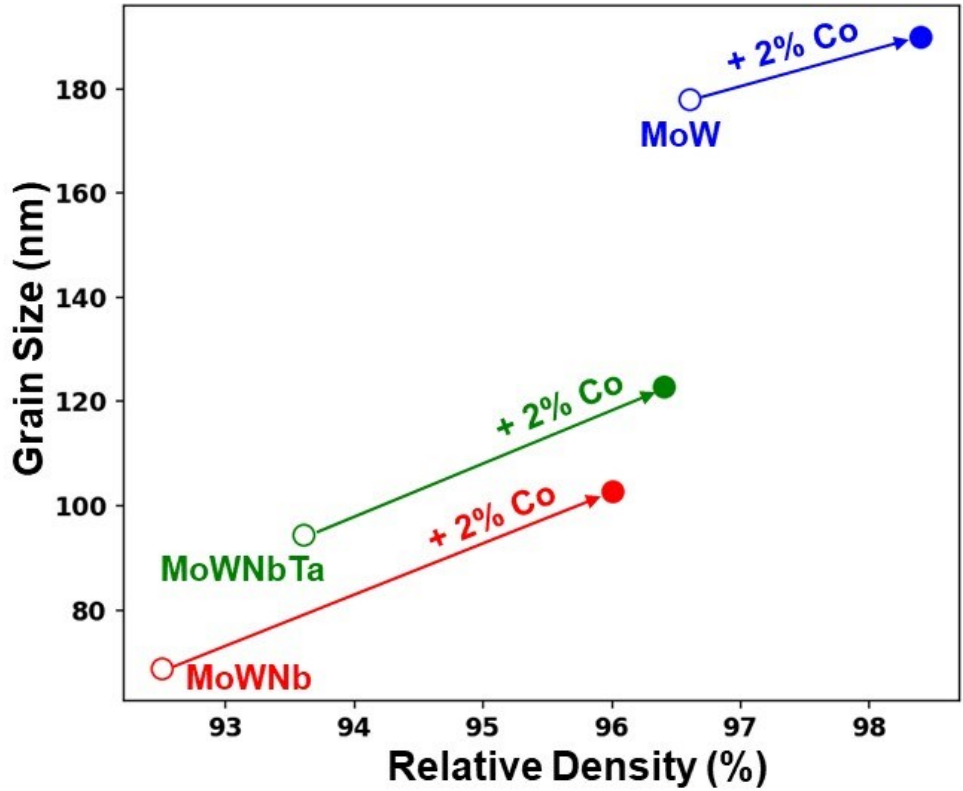

**Figure 3** Grain size as a function of relative density for as-sintered specimens of six compositions: equimolar MoW, MoWNb, and MoWNbTa, each without and with 2 at.% Co addition. The Co addition increased the relative densities by approximately 2–4 %, accompanied by moderate increases in grain sizes in the sintered specimens, likely due to both enhanced GB diffusion and reduced porosity.

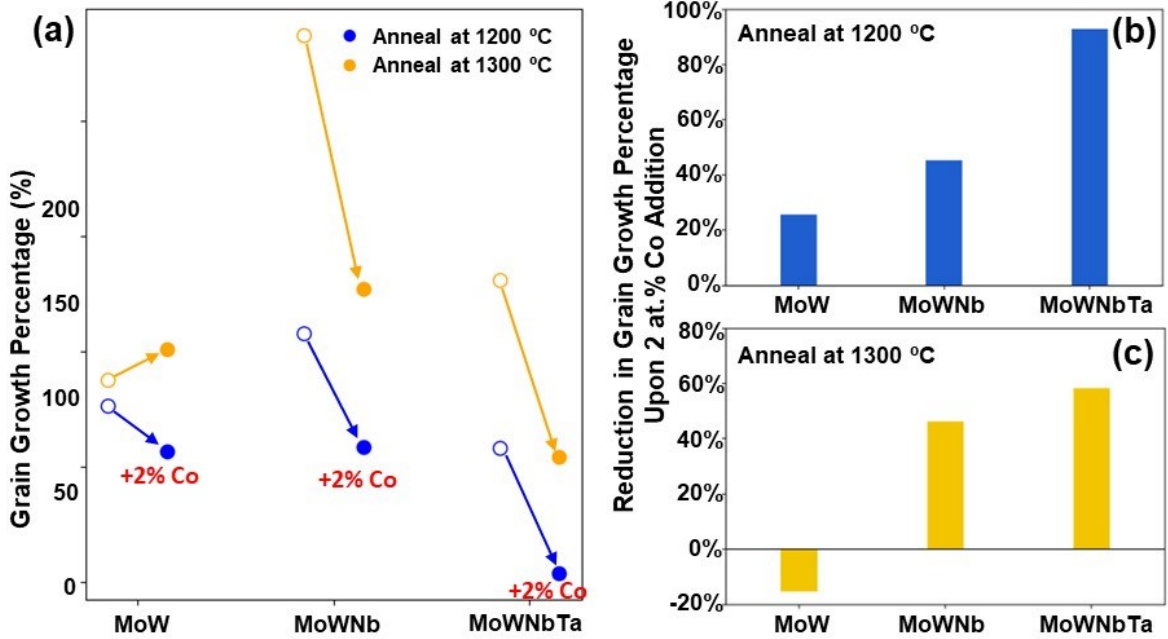

**Figure 4** (a) Grain growth percentages for all six compositions after isothermal annealing at 1200 °C and 1300 °C for 5 h. Results for undoped specimens are shown by open circles while results for specimens with 2 at.% Co addition are shown by the solid circles and labeled as "+2% Co," and the changes upon Co addition are indicated by arrows. Reduction in grain growth percentage upon 2 at.% Co addition for annealing at (b) 1200 °C and (c) 1300 °C. The results indicate that Co addition more effectively suppresses grain growth with increasing number of principal elements, following the order MoW → MoWNb → MoWNbTa, consistent with the hypothesized Type II HEGB effect [28,29].

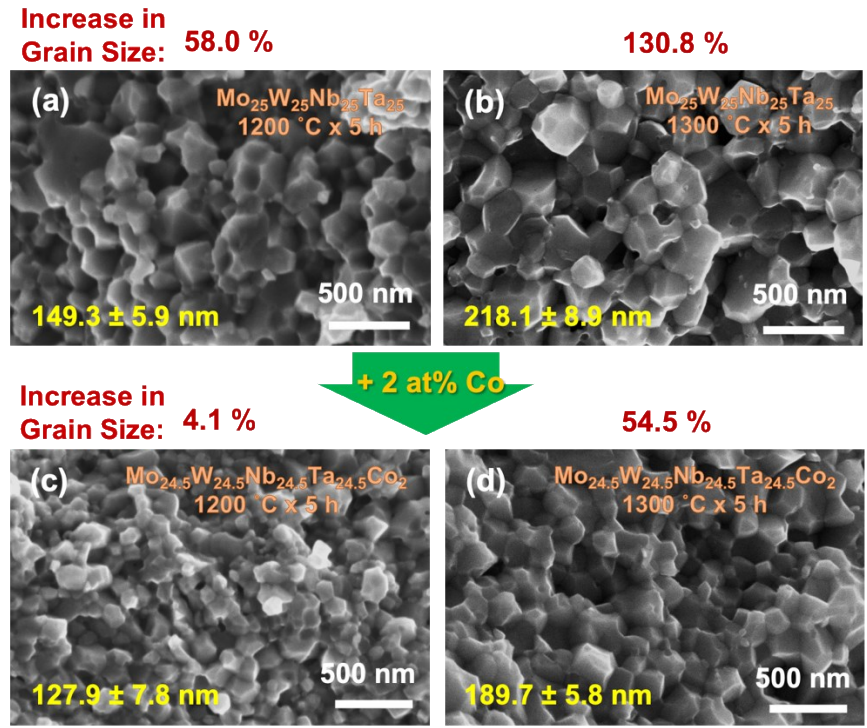

**Figure 5** Scanning electron microscopy (SEM) images of fractured surfaces of (a, b) $Mo_{25}W_{25}Nb_{25}Ta_{25}$ (*i.e.*, MoNbTaW) and (c, d) $Mo_{24.5}W_{24.5}Nb_{24.5}Ta_{24.5}Co_2$ (*i.e.*, MoNbTaW + 2 at.% Co) specimens after annealing at (a, c) 1200 °C and (b, d) 1300 °C for 5 h, respectively. The corresponding as-sintered microstructures are shown in Figure 2. The addition of Co reduced the extent of grain growth after annealing, despite the higher densities of the doped specimens. The percentage increases in grain size after 5 h of isothermal annealing are indicated for each case.

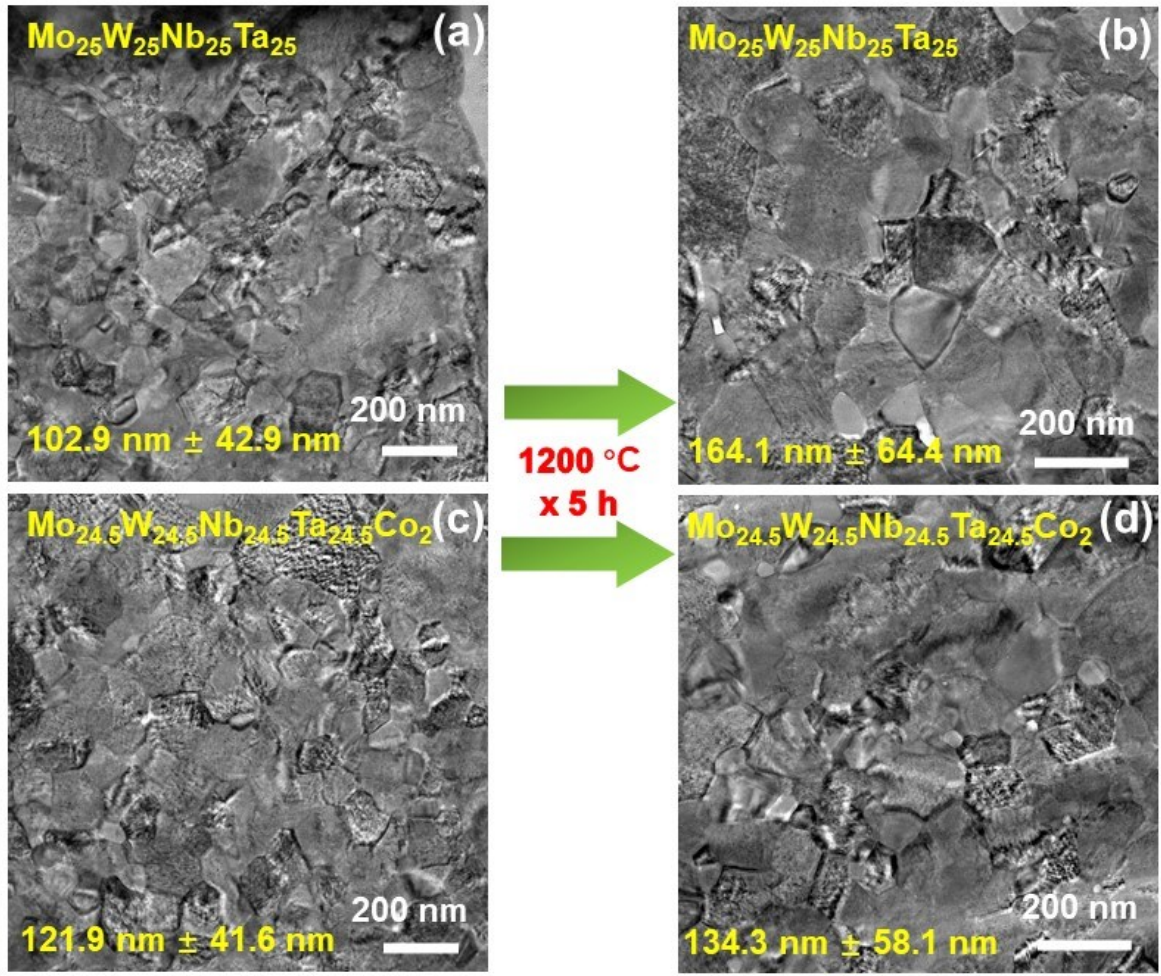

**Figure 6** Transmission electron microscopy (TEM) bright-field images of (a, b) $Mo_{25}W_{25}Nb_{25}Ta_{25}$ (*i.e.*, MoNbTaW) and (c, d) $Mo_{24.5}W_{24.5}Nb_{24.5}Ta_{24.5}Co_2$ (*i.e.*, MoNbTaW + 2 at.% Co) specimens before (a, c) and after (b, d) annealing at 1200 °C for 5 h. The SEM- and TEM-measured grain sizes are compared in Supplementary Table S1, which are consistent each other.

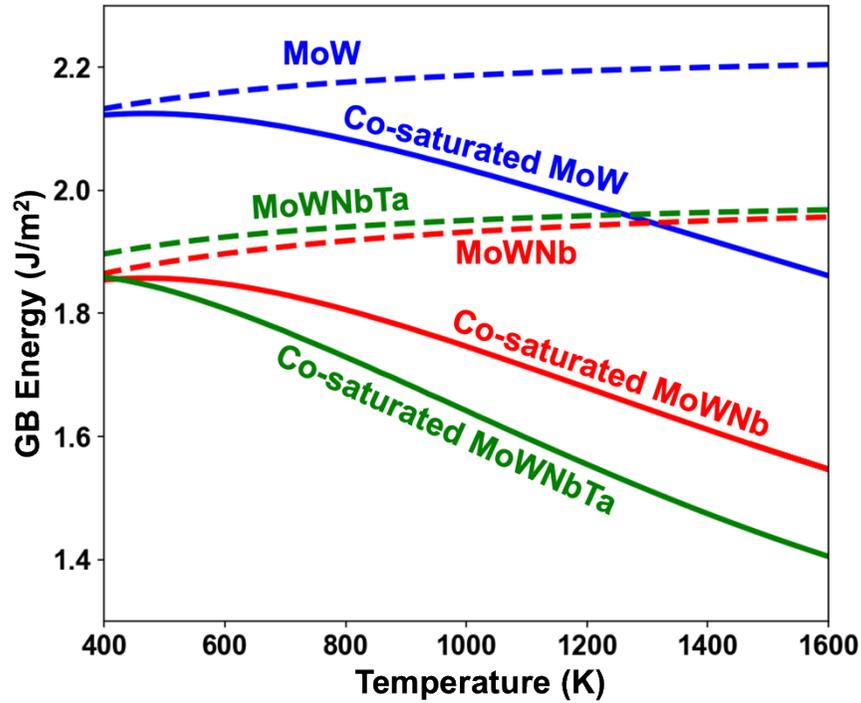

**Figure 7** Computed GB energy ($\gamma_{GB}$) as a function of temperature for undoped (dashed lines) and Co-saturated (solid lines) MoW, MoWNb, and MoWNbTa alloys. The bulk compositions of the Co-saturated BCC alloys are temperature dependent and follow the multicomponent solvus lines estimated by Equation (6), assuming a 2 at.% Co addition in equimolar RCCAs. The calculations were performed using an analytical model recently proposed by Luo [29], with model parameters obtained from the Materials Project (Supplementary Table S3).

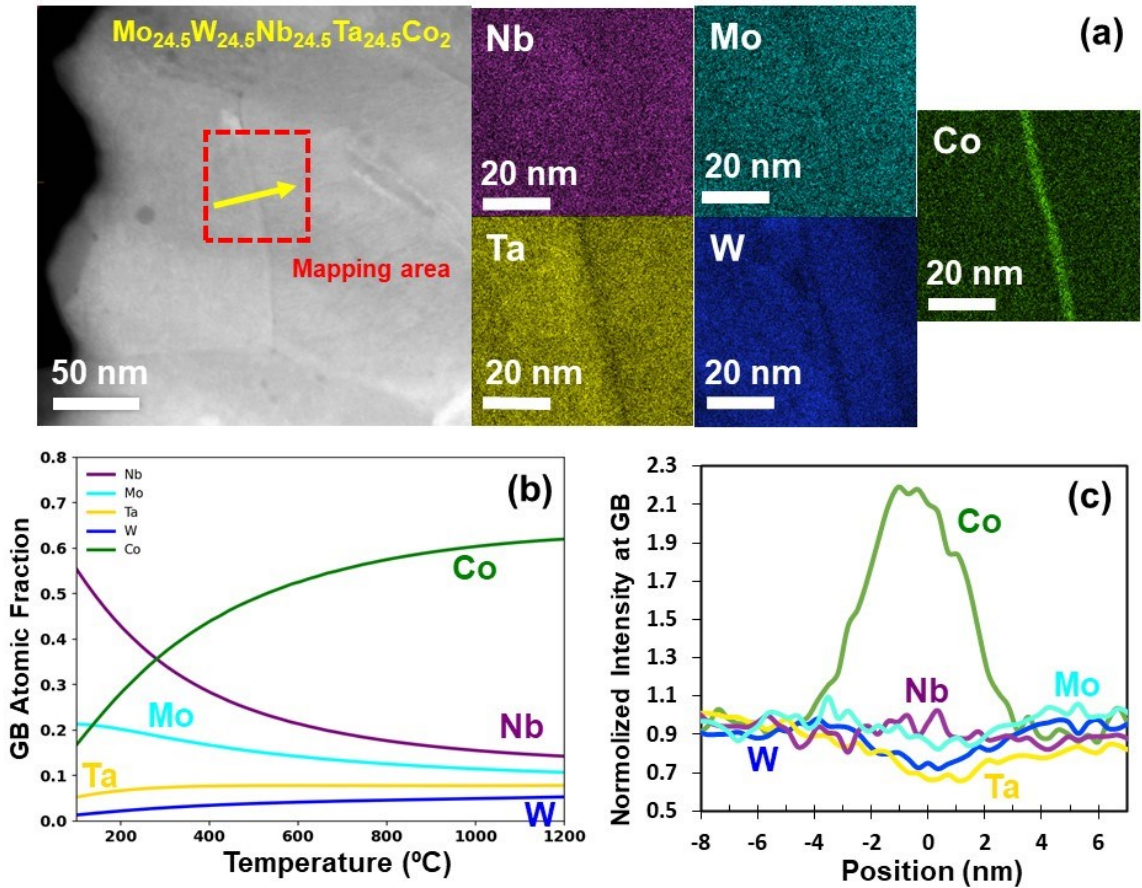

**Figure 8** (a) STEM image and EDS elemental maps of Mo$_{24.5}$W$_{24.5}$Nb$_{24.5}$Ta$_{24.5}$Co$_2$ after annealing at 1200 °C for 5 h. (b) Grain-boundary (GB) atomic fraction of each element predicted by the simplified segregation model. (c) Normalized EDS elemental intensity at the GB ($X_i^{GB}/X_i^{Bulk}$) as a function of position, obtained by integrating the EDS maps along the direction perpendicular to the GB plane located at x = 0. The data reveal strong Co segregation and mild Ta and W depletion at the GB, in agreement with the model prediction shown in panel (b).